\documentclass[12pt]{article}
\usepackage{amssymb,slashed,latexsym,ifpdf,amsmath}
\newcommand{\ft}[2]{{\frac{#1}{#2}}}

\def\rme{{\rm e}}
\def\rmi{{\rm i}}

\def\K{K\"{a}hler}
\newsavebox{\uuunit}
\sbox{\uuunit}
    {\setlength{\unitlength}{0.825em}
     \begin{picture}(0.6,0.7)
        \thinlines
        \put(0,0){\line(1,0){0.5}}
        \put(0.15,0){\line(0,1){0.7}}
        \put(0.35,0){\line(0,1){0.8}}
       \multiput(0.3,0.8)(-0.04,-0.02){12}{\rule{0.5pt}{0.5pt}}
     \end {picture}}

\newcommand{\hc}{{\rm H.c.}}
\newcommand{\bbox}{\lower.2ex\hbox{$\Box$}}
\newcommand{\Nn}{N}
\csname @addtoreset\endcsname{equation}{section}
\ifx\pdfoutput\undefined
   \pdffalse
   \usepackage{cite}
 \else
   \pdfoutput=1
   \pdftrue
  \usepackage[pdftex]{hyperref}
\fi
\newcommand{\nmult}{^1}
\newcommand{\bnmult}{^1}
\newcommand{\onen}{1}

\newcommand{\rf}[1]{(\ref{#1})}

\def\aD3{{\overline {\rm D3}}}
\def\be{\begin{equation}}
\def\ee{\end{equation}}
\def\ba{\begin{array}}
\def\ea{\end{array}}
\begin{document}

 %%%%%%%%%%%%%%%%%%%%%%%%%%%%%%%%%%%%%%%%%%%%%%%%%%%%%%%%%%%
\begin{titlepage}
\phantom{.}
\vspace{1.5cm}
\hskip 1cm
\vskip 0.5cm
\begin{center}
\baselineskip=16pt

%%%%%%%%%%%%%
{\Huge {\bf Pure de Sitter supergravity}}

\

\

\

 { \large  \bf Eric A. Bergshoeff$^1$},  {\large  \bf Daniel Z.  Freedman$^{2,3}$},   {\large \bf Renata Kallosh$^2$},\\

\vskip 0.5 cm

   {\large  \bf Antoine Van Proeyen$^4$} \vskip 0.8cm
{\small\sl\noindent
$^1$ Van Swinderen Institute for Particle Physics and Gravity,\\
University of Groningen, Nijenborgh 4, 9747 AG Groningen, Netherlands\\\smallskip
$^2$ SITP and Department of Physics, Stanford University, Stanford, California
94305 USA \\\smallskip
$^3$ Center for Theoretical Physics and Department of Mathematics, Massachusetts Institute of Technology, Cambridge, Massachusetts 02139, USA \\\smallskip
$^4$ KU Leuven, Institute for Theoretical Physics,\\ Celestijnenlaan 200D, B-3001 Leuven,
Belgium}

\end{center}

%%%%%%%%%%%%%%%%%%%%%%

\vskip 2.5cm
\begin{center}
{\bf Abstract}
\end{center}
{\small Using superconformal methods we derive  an explicit de Sitter supergravity action invariant under spontaneously broken local ${\cal N}=1$ supersymmetry. The supergravity multiplet interacts  with a  nilpotent Goldstino multiplet. We present a complete locally supersymmetric action including the graviton and the fermionic fields, gravitino and Goldstino,   no scalars.  In the global limit when the supergravity multiplet decouples, our action reproduces the  Volkov-Akulov theory. In the unitary gauge where the Goldstino vanishes we recover pure supergravity with the positive cosmological constant.  The classical equations of motion,  with all fermions vanishing,  have a maximally symmetric solution:  de Sitter space.  }\vspace{2mm} \vfill \hrule width 3.cm
{\footnotesize \noindent e.a.bergshoeff@rug.nl, dzf@math.mit.edu, kallosh@stanford.edu,
antoine.vanproeyen@fys.kuleuven.be }
\end{titlepage}
\addtocounter{page}{1}
 %\tableofcontents{}
\newpage
%%%%%%%%%%%%%%%%

\section{Introduction}

The cosmological constant is known to be negative or zero in pure supergravity, if there are no scalar fields  \cite{Townsend:1977qa}. Pure supergravity with a positive cosmological constant without scalars was not previously known.
In this paper we present the locally ${\cal N}=1$ supersymmetric action and transformation rules of such a theory. De Sitter space is a homogeneous solution of the bosonic equations of motion.  Supersymmetry is spontaneously broken, so there is no conflict with no-go theorems that prohibit linearly realized supersymmetry \cite{Pilch:1984aw}.\,\footnote{Note that there exist ${\cal N}$-extended  de Sitter superalgebras for even ${\cal N}$ but they have a noncompact $R$-symmetry group and therefore do not allow unitary representations.}

The main motivation for this work is an increasing amount of observational evidence for an accelerating Universe where a positive cosmological constant is a good fit to data. The next step toward a better understanding of dark energy is not expected before the  ESA space mission Euclid launches in 2020.
It is therefore desirable to find a simple version  of de Sitter supergravity as a natural source for the positive cosmological constant.

The Kachru, Kallosh, Linde and Trivedi (KKLT) uplifting procedure for constructing de Sitter (dS) vacua in string theory  was proposed in \cite{Kachru:2003aw}. It  was recently updated to the status of a manifestly supersymmetric uplifting using the $ \aD3 $-brane on top of an O3-plane at the bottom of a warped throat \cite{Bergshoeff:2015jxa,Kallosh:2015nia}. It corresponds to a globally supersymmetric  Volkov-Akulov (VA) Goldstino theory \cite{Volkov:1973ix} coupled to a supergravity background. The global supersymmetry is realized nonlinearly. This recent development indicates that
a scalar independent de Sitter supergravity might exist.
Another indication of the existence of such a supergravity was presented in \cite{Deser:1977uq}, where the proposal  to couple the VA Goldstino theory \cite{Volkov:1973ix} to supergravity was made. However, a complete action and transformation rules that describe this coupling have never been presented. The supersymmetric coupling of the gravitino and Goldstino in $D=10$ at the
quadratic level in fermions was studied in \cite{Dudas:2000nv,Pradisi:2001yv}.
The curved superspace formulation of the VA Goldstino theory was studied soon after the discovery of this theory; see for example a review paper
\cite{Volkov:1994vg} or an application of the constrained superfield formalism in superspace in \cite{Samuel:1982uh}. The relation between the superspace approach and nonlinearly realized supersymmetries  was investigated  in \cite{Ivanov:1989bh}.

All earlier  theories were  not yet developed to the level of  a component supergravity action with spontaneously broken local supersymmetry, generalizing the globally supersymmetric VA model. To construct such an action is the purpose of our paper. We will do this by decoding the superconformal action underlying dS supergravity, proposed in \cite{Ferrara:2014kva}. Such a decoding procedure, in addition to  a standard gauge-fixing of local Weyl, $R$-symmetry and special  supersymmetry requires an elimination of the auxiliary field $F$ of the Goldstino multiplet from the  action which has a non-Gaussian dependence
on $F$.

The important step for our ability to derive  the complete action of a pure dS supergravity is the observation made in \cite{rocek,Komargodski:2009rz} that VA theory can be described using a chiral superfield  $S(x, \theta)=X +\sqrt 2\,  \theta \chi + \theta^2 F $ of global ${\cal N}=1$ supersymmetry that satisfies the nilpotent constraint $S^2(x, \theta)=0$.  The constraint sets  $X= \bar\chi P_L\chi/2 F$
and thus eliminates the would-be  fundamental scalar partner of the Goldstino $\chi$. The Komargodski--Seiberg(KS) model constructed in this way \cite{Komargodski:2009rz} is equivalent to the original VA geometric model.   That model with the action $\det E$, where  $E$ is a supersymmetric 1-form, is related to the model of \cite{Komargodski:2009rz}  by the nonlinear change of variables presented in \cite{Kuzenko:2010ef}. The fact that  $X$ is Grassmann valued so that $X P_L\chi =0$ greatly simplifies the construction of \cite{Komargodski:2009rz} and of our locally supersymmetric extension.

The superconformal approach to pure de Sitter supergravity suggested in   \cite{Ferrara:2014kva} is the following:  The model at the superconformal level contains  the chiral compensating multiplet    $\{X^0, \chi^0, F^0\}$; a chiral Goldstino multiplet  $\{ X^1  ,\chi^1 ,F^1  \}$;
and a Lagrange multiplier multiplet $\{ \Lambda, \chi^\Lambda, F^{\Lambda} \} $, interacting with the Weyl gravitational multiplet. The action  is
 \begin{equation}
{\cal L} =  [\Nn (X,\bar X)]_D + [\mathcal{W}(X)]_F + \left[\Lambda  (X\nmult )^2\right] _F\,.
\label{symbL1}
\end{equation}
where the notation of Chapter 16 of \cite{Freedman:2012zz}  is used\footnote{The superconformal action  \rf{symbL1} without the Lagrange multiplier superfield  $\Lambda$ was studied in application to inflation and in de Sitter background in \cite{Kallosh:2000ve}.
} and all three chiral supermultiplets are unconstrained.   All supersymmetries in  \rf{symbL1} are linearly realized and manifest.
Models of this type differ from
the generic models  in \cite{Freedman:2012zz}, and in  other textbooks,  in that the \K\, manifold of the embedding space $\Nn (X,\bar X)$ does not depend on the superfield $\Lambda$ but does depend on $X^I$, $I=0,1$. Therefore the equation of motion for $\Lambda$ is algebraic and can be solved producing the superfield constraint $(X\nmult )^2=0$. This in turn leads to a nongeneric supergravity: the elimination of the auxiliary  field $F^1$ requires a more  complicated procedure since its algebraic equation of motion contains both positive and negative powers of $F^1$,  the latter
due to  the  relation $X^1= \bar\chi^1 P_L\chi^1/2 F^1$ which arises as the solution of the constraint.
%constraint resolving the nilpotency condition $(X\nmult )^2=0$.
Therefore, the  knowledge of the \K\, potential $K$ and the superpotential $W$ at the supergravity level is not sufficient in presence of the nilpotent Goldstino multiplet to produce the full fermionic  action.\footnote{See Eqs. (2.4) - (2.6) in \cite{Antoniadis:2014oya} where the first supergravity model of this kind was presented. The fermion terms in this reference are incomplete, which gives an example of a supergravity where $K$ and $W$ are not sufficient for the determination of the complete action; only the bosonic part can be deduced from $K$ and $W$. }

At the superconformal level, the dynamics of our pure dS supergravity model  is specified by a quadratic K\"ahler potential and cubic superpotential:
\begin{equation}
  \Nn=\eta _{IJ}X^I\bar X^J=-X^0\bar X^0 +X\nmult   \bar X\bnmult \,,\qquad \mathcal{W}= a\left(\frac{X^0}{\sqrt{3}}\right)^3 + b\left(\frac{X^0}{\sqrt{3}}\right)^2 X\nmult \,.
 \label{NWchoice}
\end{equation}
After the superfield constraint $(X\nmult )^2=0$  is implemented, the last term in the action \rf{symbL1} vanishes.  The parameters  $a,~b$ are dimensionless as they must be in a conformal theory.  One passes to the physical form of the theory by fixing the conformal gauge using $X^0= \sqrt{3}/{\kappa}$, thus introducing Newton's constant $\kappa^2=8\pi G= M_{Pl}^{-2}.$  It is then convenient\footnote{In cosmological applications one often works with Planck units, $M_{Pl}= \kappa^{-1}=1$, however, here we would like to study also the flat space limit. Therefore we keep $\kappa$ consistently, in agreement with \cite{Freedman:2012zz}.} to  redefine our parameters as follows:  $a = \kappa \, m  $ and $  b= \kappa^2 \, f $.  The new parameters $m$ and $f$ have mass dimension 1 and  2, respectively, and we take them to be real.\footnote{In Sec. \ref{ss:derivation} and the Appendix we give the formulas for complex $a$ and $b$ (or $m$ and $f$). Then $b= \kappa^2 \bar f$. One can take these two parameters real and positive after chiral rotations of the fields. If $a=|a|\rme^{\rmi\theta _a}$ and $b=|b|\rme^{\rmi\theta _b}$, the phases are removed by replacing $P_L\psi _\mu$ by $P_L\psi _{\mu} \rme^{\rmi\theta _a/2}$, $P_L\chi$  by $P_L\chi \rme^{\rmi(\theta _a/2-\theta _b)}$,
% $X=X^1$ by $X^1\rme^{\rmi(\theta _a-\theta _b)}$ and $F=F^1$ by $F^1\rme^{-\rmi\theta _b}$
 and corresponding rotations on the composite expressions $X=X^1$ and $F^I$.}
 The  cosmological constant ${\bf \Lambda}$ and the Lagrangian mass term of the gravitino are
\begin{equation}
{\bf \Lambda} = {f^2- 3\,  {m^2\,  M_{Pl}^2} },\qquad\quad  L_m =  \frac{m}{2\kappa^2 } { \bar\psi}_\mu \gamma^{\mu\nu} \psi_\nu \,,
\end{equation}
where $\psi_\mu$ has dimension 1/2.

The physics of the model depends on the relation between these quantities.  When $m=0,~f\ne 0$, we have the pure de Sitter model with nonlinearly realized supersymmetry discussed above.
When $m \ne 0,~ f=0$, which requires  that the fermion of the nilpotent multiplet vanishes, $\chi^1=0$;  for consistency,  we have the basic anti--de Sitter supergravity theory with linearly realized supersymmetry \cite{Townsend:1977qa,Deser:1977uq}.   In all other cases there is nonlinearly realized supersymmetry,   and the sign of ${\bf \Lambda}$ determines whether the homogeneous bosonic geometry is de Sitter, Minkowski, or anti--de Sitter spacetime. Nonlinearly realized supersymmetry (essentially the same as spontaneous breaking) means that the vacuum expectation value of the SUSY transform of the Goldstino field $\chi$ does not vanish, $ \langle \delta\chi\rangle \ne 0$.

In Sec. \ref{ss:resultdS} of the paper we present the main result, the novel pure dS supergravity action and its local  supersymmetry. In Sec. \ref{ss:derivation} we explain the main logical steps in the derivation of the supergravity theory from the superconformal model in \cite{Ferrara:2014kva}, with the details given in the Appendix. In Sec. \ref{ss:features} we study features of dS supergravity. We perform the limit of our new supergravity theory to  flat spacetime, where  fields of the gravity multiplet are decoupled and $m\rightarrow 0$. We show how the VA theory is recovered via its KS version. In the same section we look at the possible gauge-fixing of the local supersymmetry. In the unitary gauge with $\chi^1=0$ the gravitino wave operator in Euclidean
signature has no zero modes.
In Sec. \ref{ss:discussion} we point out that the assumption of the mere existence of the nilpotent Goldstino multiplet signifies a natural unavoidable
spontaneous supersymmetry breaking, without the need for engineering, as e.g in the O'Raifeartaigh type models.
 Finally we note that the VA theory, when embedded in supergravity, leads to a positive cosmological constant term ${\cal L}_{SG} = -\sqrt {- \det g} \, f^2$. Without coupling to gravity and gravitino, without local supergravity, the  vacuum energy term in the VA action ${\cal L}_{\rm VA} = - f^2+...$ is a hint but not a reliable origin of the  dark energy/cosmological constant; now in the context of  dS supergravity it is a cosmological constant!

\section{Pure dS \texorpdfstring{${\cal N}=1$}{N=1} supergravity action and its local supersymmetry}
\label{ss:resultdS}

The action invariant under spontaneously broken local supersymmetry is given by the following expression
\begin{eqnarray}
 e^{-1}{\cal L} & = &\frac{1}{2\kappa^2} \left[ R(\omega (e )) -\bar \psi _\mu \gamma ^{\mu \nu \rho } D^{(0)}_\nu \psi _\rho
 +{\cal L}_{\rm SG,torsion} \right] +3 \frac{m^2}{\kappa^2} - f^2 \nonumber\\
&&+\frac{f}{\sqrt{2}} \bar \psi_\mu \gamma
  ^\mu   \chi+\frac{m}{2\kappa^2}  \bar \psi _{\mu } \gamma ^{\mu \nu }\psi _{\nu } +\frac{\kappa^2}{24} \chi ^2\bar \chi ^2\nonumber\\
 &&-\ft12\bar \chi  \slashed{D}^{(0)}\chi -\ft{1}{32}\rmi\,e^{-1}\varepsilon ^{\mu \nu \rho \sigma }\bar \psi _\mu \gamma _\nu \psi _\rho \bar \chi \gamma _*\gamma _\sigma \chi-\ft12\bar \psi _\mu P_R\chi \bar \psi ^\mu P_L\chi\nonumber\\
&& +\frac{\bar  \chi^2}{2 f}A \frac{\chi^2}{2 f} -\left(\frac{ \chi^2}{2 f} \bar B+\frac{\bar \chi^2}{2 f} B\right) - \frac{\chi^2\bar \chi^2}{16 f^4}\left(\frac{\bbox\chi^2}{f}-2B\right)\left(\frac{\bbox\bar \chi^2}{f}-2\bar B\right)\,,\label{finalactionf1}
\end{eqnarray}
where
\begin{eqnarray}
\chi ^2&\equiv& \bar \chi P_L\chi \,,\qquad   D^{(0)}_\mu    = \partial _\mu +\ft14 \omega _\mu {}^{ab}(e )\gamma _{ab}\,,\nonumber\\
  {\cal L}_{\rm SG,torsion}&=&-\ft{1}{16}\left[
(\bar{\psi}^\rho\gamma^\mu\psi^\nu) ( \bar{\psi}_\rho\gamma_\mu\psi_\nu
+2 \bar{\psi}_\rho\gamma_\nu\psi_\mu) - 4 (\bar{\psi}_\mu
\gamma\cdot\psi)(\bar{\psi}^\mu \gamma\cdot\psi)\right]\nonumber\\
&&-e^{-1}\partial _\mu \left(e\bar \psi \cdot \gamma \psi ^\mu \right)\,.
\label{covder0}
\end{eqnarray}
\begin{equation}
  A= \bbox + \rmi t^\mu \partial _\mu + \ft12\rmi e^{-1}\partial _\mu (e\,t^\mu) + r\,, \qquad
  \bbox =\frac{1}{\sqrt{g}}\partial _\mu \sqrt{g}g^{\mu\nu}\partial _\nu\,,
 \label{formA}
\end{equation}
\begin{eqnarray}
t^\mu &=&  \ft14 {\rmi} {\bar \psi}_\nu \gamma_* \gamma^{\nu\rho\mu}\psi_\rho\,,\quad
r =-\ft16 \left[R(\omega (e )) -\bar \psi _\mu \gamma ^{\mu \nu \rho } D^{(0)}_\nu \psi _\rho
    + {\cal L}_{\rm SG,torsion}
   -8\kappa ^2\,f^2\right] \,,\nonumber\\
 B  &=& \frac{1}{\sqrt{2}}\left[-e^{-1}\partial _\mu \left(e\bar \psi _\nu \gamma ^\mu \gamma ^\nu P_L  \chi  \right)-\ft23\bar \chi P_L \gamma ^{\mu \nu }D _\mu \psi _\nu  \right]
 + f
\Bigl(2 m +   % corrected
\frac{1}{2} {\bar\psi}_\mu\gamma^{\mu\nu}P_L\psi_\nu\Bigr)\,,\label{formB}
\end{eqnarray}
\begin{equation}
  D_\mu \psi _\nu=\left(\partial _\mu +\ft14 \omega _\mu {}^{ab}(e,\psi )\gamma _{ab}\right) \psi _\nu\,.
 \label{Dmupsinu}
\end{equation}
 The auxiliary fields $F$ and $A_\mu$ of the supergravity multiplet were eliminated by their algebraic equations of motion.
The cosmological constant  in the first line is ${\bf \Lambda}=  f^2- 3 \frac{m^2}{\kappa^2} $. In the second line the gravitino couples to $f\gamma^\mu\chi$ which is the linear part of the supercurrent of the VA theory; nonlinear corrections are contained in $B$ in (\ref{formB}).
There is also a  quadratic gravitino masslike term and a quartic $\chi ^2\bar \chi ^2$ originating from elimination of $A_\mu$. The third line of the action includes a Goldstino kinetic term and quartic fermion  interactions. The fourth line presents nonlinear  Goldstino terms.

The supersymmetry transformations of the fields $\chi$ and $e^a_\mu,~\psi_\mu$ can be obtained from, respectively,  (16.33), (16.45), and (16.47) of \cite{Freedman:2012zz}.  For the fields of the gravity multiplet we
have
\begin{eqnarray}
\delta e^a_\mu &=& \frac12 \bar\epsilon \gamma^a\psi_\mu \ ,\\
\delta P_L\psi_\mu &=& P_L\bigg(\partial_\mu +\frac14 \omega_{\mu ab}(e,\psi)\gamma^{ab}-\frac32iA_\mu +\frac12i\gamma_\mu \slashed{A} +
\frac{\kappa}{2\sqrt3}\gamma_\mu \bar F^0 \bigg)\epsilon\label{deltpsi}\,.
\end{eqnarray}
  with
\begin{equation} % \kappa factors corrected
  F^0=\overline{{\cal W}}_0= \sqrt{3}\,\frac{m}{\kappa}  +\frac{2}{\sqrt{3}} \kappa f\, X= \sqrt{3}\,\frac{m}{\kappa}- \frac{1}{\sqrt{3}}\kappa \chi ^2(1-{\cal A})\,.
 \label{Fvalue}
\end{equation}
and
\begin{eqnarray}
A_\mu= \rmi\frac{\kappa^2}{6}\Big [ (\bar X\partial _\mu X -X\partial _\mu \bar X )  -\frac{1}{2}[\sqrt{2}\bar \psi _\mu(P_L\chi \bar X -P_R  \chi  X) +\bar \chi P_L\gamma_\mu \chi ]\Big ] .
 \label{Amu}\end{eqnarray}
 Here
 \be
 X= - \frac{\chi^2}{ 2 f} (1-{\cal A}) \ ,
\label{XinA}  \ee
  \be
    {\cal A}= \frac{\bar \chi^2}{ 2 f^3} \left(A\, \frac{\chi^2}{2f} - B\right).
 \label{calA}   \ee
The local supersymmetry transformation  for the Goldstino is
%\begin{equation}
%\delta P_L \chi =-  \frac{f}{\sqrt 2} P_L \epsilon (x)  + \frac1{\sqrt 2} P_L  \Big ( \slashed{\cal D} X -  f\left[{\cal A} \left(1-3\bar{{\cal A}}-\frac{\chi^2}{2f^3}\bar B\right)\right]   \Big )\epsilon (x) + \sqrt 2 P_L X \eta \ ,
%\label{deltgold}\end{equation}
%where
%\begin{eqnarray}
%  P_L\eta &=& \frac12\rmi P_L\slashed{A}\epsilon -\frac{\kappa
%  }{2\sqrt{3}}F^0P_L\epsilon \,,\nonumber\\
%   {\cal D}_\mu  X &=&\partial _\mu  X- \rmi \, A_\mu
%  X -\frac{1}{\sqrt{2}}\bar \psi _\mu P_L\chi \, .
% \label{definitionssusy}
%\end{eqnarray}
\begin{equation}
 \delta P_L\chi =\frac1{\sqrt{2}} P_L\left[-f+(\slashed{\partial }-m)X
-  f {\cal A} \left(1-3\bar{{\cal A}}-\frac{\chi^2}{2f ^3}\bar B\right) \right]\epsilon
-\frac{1}{2}P_L\gamma ^\mu \epsilon \bar \psi _\mu P_L\chi
\,.
\label{deltgold}
\end{equation}

\section{Derivation of  pure dS supergravity}
\label{ss:derivation}
In this section we present the main steps in the derivation of dS supergravity from the underlying superconformal theory with linearly realized supersymmetry and Lagrange multiplier, as shown in Eqs. \rf{symbL1}, \rf{NWchoice}. Details are given in  the Appendix.

We will often use the notation for the physical multiplet $\{X^1, \chi^1, F^1\}\equiv \{X, \chi, F\}$. The role of the compensator multiplet $\{X^0, \chi^0, F^0\}$ is to fix the local Weyl and $R$-symmetry  via the choice $X^0=\bar X^0 = \frac{\sqrt 3}{\kappa}$ and to fix the special local supersymmetry using $\chi^0=0$. But in a superconformal setting  where equations depend covariantly on both multiplets $\{X^I, \chi^I, F^I\}, \, I=0,1$  we will use the original notation.

We first consider the component form of the Lagrange multiplier term in the action in \rf{symbL1} and solve the algebraic equations of motion for the superfield
$\{\Lambda, \chi^\Lambda, F^\Lambda\}$. The  $\Lambda(x)$ field equation  is given in \rf{fieldeqnLambda};  its solution fixes
 \begin{equation}
  X = \frac{\chi^2 }{2F }\,,
 \label{Xnsolved}
\end{equation}
provided that $F\neq 0$. The equations of motion for   $\chi^\Lambda, ~F^\Lambda $ are also satisfied without further constraints.

The detailed form of the  first two terms in the superconformal action \rf{symbL1}  is given in \rf{fullactionFform} which we then write as
\begin{eqnarray}
 {\cal L}
 & = &\eta _{IJ} \bar X^I  \left[\partial _\mu \sqrt{g}g^{\mu\nu}\partial _\nu + \rmi\,e\, t^\mu_c \partial _\mu + \ft12\rmi\,\partial _\mu (e\,t^\mu_c) + e\,r_0^c\right] X^J
  \nonumber\\
   &   & +e\, \eta _{IJ}\bar X^I\, B_c^J + e\,\eta _{IJ}X^I\bar B^J_c +e \, C_0^c+ e\,{\cal L}_{1,F}+e\,{\cal L}_{W,\rm ferm}\,,
 \label{12XbarXD}
\end{eqnarray}
(up to total derivatives). The indices $I=0,1$ and the subscript $c$ are a reminder that we are still in the superconformal setting with local conformal
symmetry (and other symmetries) unbroken:
\begin{eqnarray}
t^\mu_c &=&-2 A^\mu +\ft14 {\rmi} {\bar \psi}_\nu \gamma_\star \gamma^{\nu\rho\mu}\psi_\rho\,,\nonumber\\
r_0 ^c&=&-\ft16 R(\omega (e )) +\ft16\bar \psi _\mu \gamma ^{\mu \nu \rho } D^{(0)}_\nu \psi _\rho
  -A^\mu A_\mu -\ft16{\cal L}_{\rm SG,torsion} \,,\nonumber\\
B^I_c &=& \frac{1}{\sqrt{2}}\left[-e^{-1}\partial _\mu \left(e\bar \psi _\nu \gamma ^\mu \gamma ^\nu P_L  \chi^I\right)-\ft23\bar \chi^I P_L \gamma ^{\mu \nu }D _\mu \psi _\nu +\rmi A^\mu \bar \psi _\mu P_L\chi^I\right]\,, \nonumber\\
  C_0^c&=&\eta _{IJ} \left(  -\ft12\bar \chi^I  \slashed{D}^{(0)}\chi^J+\ft14\rmi \bar \chi^I \gamma _*\gamma ^\mu \chi^J A_\mu \right.\nonumber\\
 &&\phantom{\eta _{IJ}\ }  \left.-\ft{1}{32}\rmi\,e^{-1}\varepsilon ^{\mu \nu \rho \sigma }\bar \psi _\mu \gamma _\nu \psi _\rho \bar \chi^I \gamma _*\gamma _\sigma \chi^J-\ft12\bar \psi _\mu P_R\chi^I \bar \psi ^\mu P_L\chi^J \right) \,,\nonumber\\
 {\cal L}_{1,F} & = & \eta_{IJ}F^I\bar F^J+{\cal W}_IF^I +\overline{{\cal W}}_{\bar I}\bar  F^I\,,\nonumber\\
 {\cal L}_{W,\rm ferm}&=&-\ft12{\cal  W}_{IJ}\bar \chi^{I}P_L\chi ^{J}+\frac{1}{\sqrt{2}}\bar \psi_\mu \gamma
  ^\mu {\cal  W}_I P_L \chi^I+\ft12 \bar \psi _{\mu }P_R \gamma ^{\mu \nu }\psi _{\nu }\,{\cal W} +\hc \,,
 \label{12tr0C0}
\end{eqnarray}
where $ D^{(0)}_\mu$ and
$
  {\cal L}_{\rm SG,torsion}$   are defined in \rf{covder0}  and $
D_\mu \psi _\nu $ in \rf{Dmupsinu}.

The nilpotent fields $ X$ and $\bar X$ can appear in the Lagrangian
\rf{12XbarXD} either linearly or as the bilinear  $X\bar X$. Thus we  look for a new form of ${\cal L}$ in which this behavior is manifest. This form is
\begin{equation}
e^{-1}{\cal L}(X,F) = (F+\bar {\cal W}_1)(\bar F + {\cal W}_1) - \bar {\cal W}_1  {\cal W}_1  + \bar X\, A_c\, X + X\bar B_c + B_c\bar X +C_c\,.
\label{actionSC}
\end{equation}
Several  simplifications based on the superconformal properties of the
equations of motion
were required to derive this form, as explained in Appendix~\ref{ss:details}.

The main difference between dS supergravity and standard supergravities is now clear.  In a generic theory the auxiliaries $F^I$ appear as
\be\label{generic}
\eta_{IJ}F^I\bar F^J+{\cal W}_IF^I +\overline{{\cal W}}_{\bar I}\bar  F^I.
\ee
This behavior applies to $F^0$ in our theory, and this allows us to  eliminate it via Gaussian integration; we give details and the forms of the coefficients in (\ref{actionSC}) in Appendix~\ref{ss:elimauxf}.

The auxiliary field $A_\mu$ is also  eliminated in this way (see Appendix \ref{ss:elimAmu}); its on-shell value, after superconformal gauge-fixing
\begin{equation}
  X^0=\bar X^0=\kappa^{-1}\sqrt{3}\,,\qquad \chi^0=0\,,
 \label{sconfgf}
\end{equation}
is given in \rf{Amu}.
The Grassmann properties of $X,~\bar X$  imply that on-shell effects of $A_\mu$ are far simpler than in a generic supergravity. Thus $A_\mu$ vanishes in $B^I_c$ %,~
above, and the quadratic $A^\mu A_\mu$ term in $r^c_0$ with the term in $C_0^c$ produces the quartic $\chi^2\bar\chi^2$ in the second line of \rf{finalactionf1}.

The action \rf{actionSC} reduces to the form
\begin{equation}
e^{-1}{\cal L}=  (F+f)(\bar F + \bar f)% - \bar f  f 
 + \bar X\, A\, X + X\bar B + B\bar X +C\,,
\label{actionS}
\end{equation}
where, with (\ref{NWchoice}) and (\ref{sconfgf}),
\begin{equation}
  f=\overline{{\cal W}}_1=\kappa ^{-2}\bar b\,,
 \label{finb}
\end{equation}
and, for $f$ and $m$ real, $A$ and $B$ are the expressions in (\ref{formA}) and  (\ref{formB}), and $C$ is given in (\ref{formC}).

The elimination of $F^1=F$ is a more complicated matter because the generic form no longer holds.
To see this, one
substitutes $ X= \frac{\chi^2 }{2F}$ in \rf{actionS} to obtain
\begin{equation}
e^{-1}{\cal L}= (F+f)(\bar F + \bar f) %- \bar f f
+  \frac{\bar \chi^2 }{2\bar F} \, A\, \frac{\chi^2 }{2F} + \frac{\chi^2 }{2F} \bar B + B \frac{\bar \chi^2 }{2\bar F}+C\,.
\label{action1}
\end{equation}
A closed form solution for the equations of motion for $F,~\bar F$ is derived in Appendix~\ref{app:solnF}.  We
%present a simplified discussion after the gauge-fixing (\ref{sconfgf})  and elimination of $A_\mu$. and we
find that the equation of motion for $F$ is solved by
\begin{equation}
  F= -  f\left[1+{\cal A} \left(1-3\bar{{\cal A}}-\frac{\chi^2}{2f^2\bar f}\bar B\right)\right]\,,
 \label{finalF}
\end{equation}
where  $X$ and  $ {\cal A}$ are given in  \rf{XinA} and \rf{calA}, respectively.
On shell, the action \rf{actionS} becomes
\begin{eqnarray}
  {\cal L}
  &=& %-f\bar f
  +\frac{\bar  \chi^2}{2\bar f}A \frac{\chi^2}{2 f} -\left(\frac{ \chi^2}{2 f} \bar B+\frac{\bar \chi^2}{2\bar f} B\right)
  +C\nonumber\\
  &&- \frac{\chi^2\bar \chi^2}{16(f\bar f)^2}\left(f^{-1}\bbox\chi^2-2B\right)\left(\bar f^{-1}\bbox\bar \chi^2-2\bar B\right)  \,.
 \label{finalactionf2}
\end{eqnarray}
 This leads to our final result in \rf{finalactionf1}.

\section{Features of dS supergravity}
\label{ss:features}
In this section we discuss several features of the dS supergravity theory we constructed in the previous two sections. In the first subsection we discuss the flat spacetime limit and show that the theory reduces to the global Volkov-Akulov theory. In a next subsection we confirm that the de Sitter solution of the theory has no Killing spinors;
i.e.~there is no residual supersymmetry. Finally, in a third subsection we gauge-fix the local supersymmetry and show that the gravitino operator in a de Sitter background is well defined.

\subsection{The flat spacetime limit.}

In the limit of the locally supersymmetric theory in which gravitational effects vanish, we expect to recover the Komargodski-Seiberg  version \cite{Komargodski:2009rz}
of the global VA theory.
This is the limit
in which the
parameters $\kappa$, $m\kappa ^{-1}$; the curvature $R$; and the fields $\psi_\mu,\, A_\mu$ all vanish.
 In this limit  the action (\ref{finalactionf1}) reduces to
\begin{eqnarray}
 {\cal L} =-f^2  -\frac12\bar \chi
  {\slashed{\partial  }}\chi  + \frac1{4f^2}\bar \chi^2\bbox \chi^2-\frac{1}{16f^6}\chi^2\bar \chi^2(\bbox \chi^2)(\bbox \bar \chi^2)\,,
 \label{LequivVA}
\end{eqnarray}
which is equivalent to Eq.~(3.6) of \cite{Komargodski:2009rz}.

It is worth noting that the global limit of the fields of the constrained Goldstino multiplet is given by
\begin{equation}\label{constcomps}
\left\{X = \frac{\chi^2}{2F}, ~\chi~, F= -f\left(1 +\frac{1}{4f^4} \bar\chi^2\bbox\chi^2 -\frac{3}{16f^8} \chi^2 \bar\chi^2\bbox  \chi^2\bbox\bar\chi^2\right)\right\}\,.
\end{equation}
These constrained components of the Goldstino multiplet in (\ref{constcomps}) transform as though they are elementary,  i.e.
\begin{eqnarray}
\delta X &=& \frac{1}{\sqrt{2}}\bar\epsilon P_L\chi\,,\\[.1truecm]
\delta\chi &=& \frac{1}{\sqrt{2}}P_L(\slashed{\partial} X + F)\epsilon\,,\label{dchiks}\\[.1truecm]
\delta F &=& \frac{1}{\sqrt{2}}\bar \epsilon\slashed{\partial}P_L\chi \label{df}\,.
\end{eqnarray}
This shows, above and beyond the call of duty,  that the constraint $X^2=0$ is compatible with supersymmetry.\footnote{Note that the transformation rule (\ref{dchiks}) is exactly the flat limit of the transformation rule (\ref{deltgold}).
This description of  the global supersymmetry of the KS model appears to be new;  an approximate form of $\delta\chi$ up to quadratic terms in $\chi$ was derived  in Eq.~(15) of \cite{Kuzenko:2010ef}.  Our formula (\ref{dchiks}) is exact; it terminates at eighth order because of the Grassmann properties. Since $F$ in (\ref{constcomps}) has been evaluated on shell, the SUSY transformation (\ref{df}) must be checked using the equation of motion for $\slashed{\partial}P_L\chi.$}

\subsection{No Killing spinors in dS}

We  assume that ${\bf  \Lambda} = f^2- 3m^2/\kappa^2 >0$.  Then the
homogenous bosonic solution of the equations of motion of the theory defined by the action (\ref{finalactionf1}) is de Sitter space with curvature tensor
\begin{equation}
R_{\mu\nu}^{ab} = (e^a_\mu e^b_\nu - e^a_\nu e^b_\mu)H^2\hskip .5truecm  \text{and} \hskip .5 truecm  H^2 = \kappa^2 {\bf  \Lambda}/3\,.
\end{equation}
It is  obvious that this solution has no residual supersymmetry. To see this one need only inspect the fermionic transformation rules  (\ref{deltpsi})  and (\ref{deltgold}). When $\psi_\mu$ and $\chi$ vanish, these rules simplify and give the conditions
\begin{eqnarray} \label{nosusyindS}
\delta\psi_\mu &=& \hat D_\mu \epsilon\equiv (\partial_\mu + \frac14 \omega_{\mu ab}\gamma^{ab} + \frac{m}{2}\gamma_\mu)\epsilon=0 \ ,\label{deltpsi2}\\[.1truecm]
\delta\chi &=&-\frac{f}{\sqrt2}\epsilon=0\,.
\end{eqnarray}
The second condition immediately tells us that there are no (nonvanishing) Killing spinors, indicating that the supersymmetry of the bosonic background is spontaneously broken.

The same conclusion follows from the integrability condition for (\ref{deltpsi2}). It may be useful to contrast this situation with the traditional Killing spinor analysis in anti--de Sitter space (see Sec 2.2.3 of \cite{Aharony:1999ti}).
The integrability condition for (\ref{deltpsi2}) is
\begin{equation}
[  \hat D_\mu , \hat D_\nu]\epsilon =(\frac14 R_{\mu\nu ab}\gamma^{ab} +\frac{m^2}{2}\gamma_{\mu\nu})\epsilon = \frac12\Big (H^2+m^2\Big )\gamma_{\mu\nu}\epsilon=0\,,
\end{equation}
which shows again that there are no nonvanishing solutions.

\subsection{Gauge-fixing  local supersymmetry and gravitino in dS}

The action (\ref{finalactionf1}) is locally supersymmetric.   We now impose the unitary gauge condition $\chi=0$ and the action becomes
\begin{equation}
 e^{-1}   {\cal L}_{\chi=0}
=\frac{1}{2\kappa ^2} \left[ R(e,\omega(e )) -\bar \psi _\mu \gamma ^{\mu \nu \rho } D^{(0)}_\nu \psi _\rho
 +{\cal L}_{\rm SG,torsion} \right] +\frac{3m^2}{\kappa^2} -f^2 +\frac{m}{2\kappa^2}\bar \psi _{\mu } \gamma ^{\mu \nu }\psi _{\nu }\,.
 \label{finalactionfFixed}
\end{equation}
In this Lagrangian, $f$ is the measure of spontaneous supersymmetry breaking.
When $f=0$ the theory reduces to the well-known  $AdS_4$ supergravity \cite{Townsend:1977qa}. The action (\ref{finalactionfFixed}) is locally supersymmetric uniquely in this case, so that the Lagrangian with $m\ne 0$, and ${\bf  \Lambda} =-3m^2/\kappa^2$ has effectively zero physical gravitino mass \cite{Deser:1977uq}.
The concept of the ``mass spectrum" in AdS space  is somewhat tricky;  see for example a discussion of this issue with regard to the gravitino in \cite{Deser:2000de}. It is suggested there that the spin 3/2 particle is massless in AdS space not when $m=0$ but whenever gauge invariance appears. In the AdS case above, the gauge symmetry in the action \rf{finalactionfFixed} appears in case that $f=0$ which means ${\bf \Lambda}= -3 m^2/\kappa^2$.

For ${\bf  \Lambda} = f^2- 3m^2/\kappa^2 >0$ we have dS supergravity with a positive cosmological constant. In this case, as long as ${\bf  \Lambda}>0$ there is no criterion to distinguish between ``Lagrangian'' mass $m$ and a more ``physical" mass. The reason is that
at  $f\neq 0$ the action in \rf{finalactionfFixed} never acquires a local supersymmetry unless the numerous Goldstino dependent terms are added to the action and it becomes the expression in  \rf{finalactionf1}. In particular the restoration of gauge invariance  requires a  coupling between $\gamma^\mu\psi_\mu$ and a Goldstino $\chi$.
 Therefore the wisdom accumulated in studies of the gravitino in AdS space, although nontrivial, cannot be applied for  dS supergravity in \rf{finalactionf1}.
Of course,  $ {\bf  \Lambda} = f^2- 3m^2/\kappa^2 >0$ describes a useful relation between the ``Lagrangian'' gravitino mass, the supersymmetry breaking scale and the cosmological constant.

We will confirm that the gravitino propagator\footnote{See \cite{Deser:2000de} for an application of gravitino propagators in dS and AdS spacetime to the problem of discontinuities in the massless limit.}
is well defined in dS space by showing that the wave operator in Euclidean signature has no zero modes.  Towards this end we consider the wave equation on
$S^4$ which is the Wick rotation of dS$_4$.  The radius of the sphere is given by $H^2 = \kappa^2 {\bf  \Lambda}/3$.  Consider now the  mode equation
\begin{eqnarray} \label{modeq}
&& \gamma ^{\mu \nu \rho } \hat D_\nu \psi _\rho =  \lambda \psi^\mu\,,\label{modeeq1}\\[.1truecm]
&& \hat D_\nu \equiv \partial_\nu +\frac14 \omega_{\nu ab}\gamma^{ab} + \frac{m}{2}\gamma_\nu\,.
\end{eqnarray}
We have moved the mass term into the definition of the traditional AdS covariant derivative \cite{Townsend:1977qa}  but note that $\omega_{\nu ab}$ is the spin connection on $S^4$.
To clarify covariance issues below we include the Christoffel connection, and thus replace
$\hat D_\nu \to \hat\nabla_\nu$.

Our goal is to show that $\lambda$ =0 is not an allowed eigenvalue, so that the wave operator is invertible.  The first step is to multiply Eq.~(\ref{modeq}) by $\gamma_\mu$, obtaining
\begin{equation}\label{good}
\gamma^{\nu\rho}\hat\nabla_\nu\psi_\rho = \frac12\lambda\gamma\cdot\psi\,.
\end{equation}
We next apply $\hat\nabla_\mu$ on (\ref{modeeq1}), obtaining
\begin{eqnarray}
&&\frac12\gamma^{\mu\nu\rho}[\hat\nabla_\mu,\hat\nabla_\nu]\psi_\rho\,=\, \frac{1}{2}\gamma^{\mu\nu\rho}[\frac14R_{\mu\nu ab}\gamma^{ab}+\frac{m^2}{2}\gamma_{\mu\nu}]\psi_\rho = \lambda\hat\nabla\cdot\psi\,, \nonumber\\[.1truecm]
&&  -\frac32(H^2 +m^2)\gamma\cdot\psi =  \lambda\hat\nabla\cdot\psi\,.\label{riccident}
\end{eqnarray}
We used $R_{\mu\nu}^{ab} = (e^a_\mu e^b_\nu - e^a_\nu e^b_\mu)H^2$ and some $\gamma$-algebra to obtain the last equality.

The original mode equation in (\ref{modeq}) can be decomposed to read
\begin{equation}
[\gamma^\mu\gamma^{\nu\rho}\hat\nabla_\nu\psi_\rho +\gamma^\nu\hat\nabla_\nu\psi^\mu -\gamma^{\nu }\hat\nabla^\mu\psi_\nu ] =  \lambda \psi^\mu\,.
\end{equation}
If we now suppose that $\psi_\mu$ is a putative zero mode, this equation simplifies markedly.
The right side vanishes and (\ref{good}) implies that the first term on the left side vanishes as well. For the third term, (\ref{riccident}) implies (assuming the non-supersymmetric case, $H^2+m^2\neq 0$ such that $\gamma \cdot \psi =0$)
\[\gamma^{\nu }\hat\nabla^\mu\psi_\nu= m\gamma ^{\nu \mu }\psi _\nu = -m\psi ^\mu \,.\]
Thus a zero mode must satisfy the simple equation
\begin{equation}\label{psieom}
(\gamma^\nu\hat\nabla_\nu-m)\psi^\mu= (\gamma^\nu\nabla_\nu + m)\psi^\mu= 0\,.
\end{equation}
The desired result can be obtained from
\begin{equation}
(\gamma^\rho\nabla_\rho+m)(\gamma^\nu\nabla_\nu-m)\psi^\mu =
(\gamma^\rho\nabla_\rho\,\gamma^\nu\nabla_\nu - m^2)\psi^\mu =0\,.
\end{equation}
Using the Ricci identity we find
\begin{equation}
[\gamma^\rho\nabla_\rho\,\gamma^\nu\nabla_\nu-m^2]\psi^\mu  =(\nabla^\nu\nabla_\nu -4H^2 -m^2) \psi^\mu=0\,.
\end{equation}
We then multiply by $\psi_\mu^* $ and integrate over the sphere to obtain
\begin{equation}
\int d^4x \sqrt{g} [ \nabla_\nu\psi_\mu^*  \nabla^\nu\psi^\mu +(4H^2+m^2)\psi_\mu^*\psi^\mu] =0\,.
\end{equation}
Since both terms are non-negative we learn that any zero mode $\psi_\mu(x)$ vanishes identically.

In the unitary gauge the local supersymmetry of the supergravity action (\ref{finalactionf1}) is broken. The validity of this gauge-fixing in a dS background for the gravitino field equations in Euclidean signature of space-time was demonstrated above: there are no zero modes. In Lorentzian
signature it means that the gravitino differential operator in dS space is invertible, by
analytic continuation from the Euclidean signature.

Much more is known about the gravitino field equations in dS space, since the gravitino is one of the important factors in cosmology. During inflation the background is near dS and during the current acceleration, if caused by a cosmological constant,  the background is a dS space. The classical gravitino equations which also follow from our gauge-fixed action \rf{finalactionfFixed} were studied in \cite{Kallosh:2000ve,Giudice:1999am}  in a Friedmann-Lema\^{\i}tre-Robertson-Walker metric as well as in a de Sitter background. The relatively simple form of the solution was obtained in the metric, conformal to flat, $ds^2= a^2(d \eta^2-d\vec x^2)$. The solution was found in the form of an expansion in momentum modes $\psi^\mu\sim \int d^3 k e^{-i \vec k\cdot \vec x} \psi^\mu_{\vec k} (\eta)$ where an explicit dependence  on the conformal time $\eta$ enters via Hankel functions depending on $|k\eta|$; see for example Eq.~(10.5) in   \cite{Kallosh:2000ve}.

\section{Discussion}
\label{ss:discussion}
In this paper we have derived the component Lagrangian and local SUSY transformation  rules describing the coupling of the nonlinear Volkov-Akulov theory \cite{Volkov:1973ix} to supergravity complete in all orders in fermions.  The two keys to our construction were

i)  the reformulation \cite{rocek,Komargodski:2009rz} of the global VA  theory in terms of a chiral superfield $X=\{X, \chi, F\}$ subject to the constraint $X^2=0$, and \\

ii) the superconformal approach to ${\cal N}=1,~ D=4$  supergravity in the form largely developed in \cite{Kallosh:2000ve}  and described in Chapter 16 of \cite{Freedman:2012zz} where earlier references from the 1980s on the superconformal approach to supergravity are also given.\\

The combination of these two methods is successful  because  the   Lagrange multiplier that enforces the constraint \cite{Ferrara:2014kva} maintains linearly realized local off-shell supersymmetry, so that
 superconformal methods govern the initial stages of the supergravity construction.

Nevertheless,  one may distinguish between generic models  of \cite{Freedman:2012zz}  in
which a model is completely specified by its holomorphic superpotential $W(z^\alpha)$ and \K\, potential $K(z^\alpha, \bar z^{\bar \alpha})$ and models with one or more constrained superfields.  In the first case the $F^\alpha$ appear in a universal quadratic fashion and they are easily eliminated.  When there are constraints the dependence on the $F^\alpha$ is still algebraic, but more complicated.  [See  (\ref{generic}), (\ref{action1}) above.]  Nevertheless, one can find $F$ in closed form because the scalar component of the constrained multiplet is quadratic in the Grassmann valued Goldstino, $X =-(1/2f)\chi^2 + \ldots\, ,$ where $f$ controls the cosmological constant.

The striking feature of our model is that it yields a pure de Sitter ${\cal N}=1$ supergravity action in which the physical fields consist of the graviton, gravitino, and Goldstino, but no scalars and no gauge multiplets.\,\footnote{In \cite{Dudas:2015eha} it has been shown that other superconformal constructions of such theories are dual to ours. In \cite{Hasegawa:2015bza} the same method as ours was used, but using a different gauge-fixing than the one given in Eq.~\eqref{sconfgf}.} Previous constructions of de Sitter supergravities require either
 a $U(1)$ gauge multiplet with Fayet-Illiopoulos coupling and a charged gravitino \cite{Freedman:1976uk} with consequent anomaly problems,  or
  O'Raifeartaigh-type   models with multiple chiral multiplets, engineered to arrange a potential positive at a local minimum.  \\
In our model de Sitter space is obtained as the homogeneous solution because
spontaneous supersymmetry breaking is unavoidable in the presence of a  fermionic Goldstino. We hope that it will be helpful for describing dark energy.

Since supersymmetry is broken in our model there are no Killing spinors.  There is a significant simplification of the action in the unitary gauge in which the Goldstino vanishes and the nonlinearities associated with it disappear. We then find a
 very simple form of supergravity with the cosmological constant ${\bf \Lambda}= f^2-3\frac{m^2}{\kappa^2}$. The equation of motion for the gravitino  in a de Sitter background has no zero modes and its solutions are known \cite{Kallosh:2000ve,Giudice:1999am}.

Another feature of our new dS supergravity model is that it reduces in the flat space limit to the VA global theory \cite{Volkov:1973ix} in the form given in \cite{Komargodski:2009rz}.  We emphasize
that the constrained components of the Goldstino multiplet transform as a conventional chiral multiplet after elimination of $F$.

There is  curious question for future work.  The elegant geometric Lagrangian of the original form of the VA theory involves the determinant of a quadratic form in the Goldstino, ${\cal L} = {\rm Det}(\delta^\mu_\nu + \bar\chi\gamma^\mu\partial_\nu\chi)$.  It is known how to couple it to a supergravity background  in the D-brane actions; however, the corresponding supersymmetry upon gauge-fixing local $\kappa$-symmetry is still a rigid supersymmetry    \cite{Bergshoeff:2015jxa,Kallosh:2015nia}. It would be useful to know whether de Sitter supergravity with local supersymmetry presented in this paper may be brought to the geometric form of the global VA theory: this
could generate  further insights into the nature of fundamental symmetries and the origin of  the  positive cosmological constant.

So far we have explicitly constructed only the complete pure dS supergravity action with local supersymmetry. More general explicit supergravity models  with constrained superfields  interacting with general matter multiplets, to all orders in fermions,  still have to be constructed. The corresponding superconformal action was already proposed in \cite{Ferrara:2014kva}, for any number of chiral multiplets $X^I$, with generic \K\, manifold and generic superpotential together with constraints on functions of chiral multiplets determined by Lagrange multipliers $\Lambda^k$:
 \begin{equation}
{\cal L} =  [N (X,\bar X)]_D + [\mathcal{W}(X)]_F + \left[\Lambda^k  A_k(X )\right] _F\,.
\label{symbLgen}
\end{equation}
None of the $\Lambda^k$ can appear in the K\"ahler potential and the $A_k(X)$ must be  algebraic functions of $X^I$. The superconformal action in \rf{symbLgen} must be decoded and the theory expressed in physical form. Extension of the procedures of this paper will be needed to investigate the physics of this more general framework.

\

In closing we note that pure and complete anti--de Sitter supergravity \cite{Townsend:1977qa} was first formulated in 1977, but the pure and complete
de Sitter supergravity is first constructed now, 38 years later. The action and its local supersymmetry transformation are presented in Sec. \ref{ss:resultdS} of this paper.

\

\section*{Acknowledgments}

We are grateful  to   J.~J.~M.~Carrasco,  K. Dasgupta, F. Denef, E. Dudas, S. Ferrara,  A. Linde, J. Louis, H. Nicolai, F. Quevedo, A. Uranga,  A. Westphal, and T. Wrase   for stimulating  discussions.  The research of D.Z.F. is supported by NSF Grant No. PHY-0967299 and the U.S. Department of Energy under cooperative research agreement DE-FG02-05ER41360.
The work of D.Z.F. and of R.K.
is supported by the SITP,   by the NSF Grant PHY-1316699 and  by the Templeton foundation grant ``Quantum Gravity Frontiers''. A.V.P. is supported in part by the FWO - Vlaanderen, Project No. G.0651.11, and in part by the Interuniversity Attraction Poles Programme initiated by the Belgian Science Policy (P7/37).
This work has been supported in part by COST Action MP1210 ``The String Theory Universe.'' E.B. and A.V.P. thank the Department of Physics of Stanford University and the Templeton foundation for the hospitality during a visit in which this work was initiated.

\appendix

\section{From superconformal action to supergravity}
\label{ss:conftoPoinc}

In order to write the $D$ terms in (\ref{symbL1}), we use the relation that for a chiral multiplet $(X,P_L\chi,F)$ of Weyl weight~1, the $D$-action can be written in the form of an $F$-action:
\begin{equation}
  [X\bar X]_D = \ft12 [X\bar F]_F\,,
 \label{DbecomesF}
\end{equation}
where $\bar F$ is the lowest component of a chiral multiplet of Weyl weight~2 since it transforms only under $P_L\epsilon $. The components of this multiplet are given in \cite[(16.36)]{Freedman:2012zz}:
\begin{equation}
  (\bar F, \slashed{\cal D}P_R\chi, \bbox^C\bar X)\,.
 \label{barFmultiplet}
\end{equation}
The explicit expression of the superconformal covariant derivative is given in \cite[(16.34)]{Freedman:2012zz} and of the superconformal d'Alembertian on a scalar field of Weyl weight~1 in \cite[(16.37)]{Freedman:2012zz}.
These steps are performed separately for the $X^0$ multiplet and for the $X\nmult $ multiplet. Therefore, we write the Lagrangian as
\begin{equation}
  {\cal L} =  [\ft12 \eta_{IJ}X^I\bar F^J]_F + [\mathcal{W}(X^I)]_F + \left[\Lambda  (X\nmult )^2\right] _F\,.
 \label{LinFformat}
\end{equation}
The superconformal $F$-type action is given in \cite[(16.35)]{Freedman:2012zz}. The first term of (\ref{LinFformat}) is identical to \cite[(16.39)]{Freedman:2012zz}, where pure ${\cal N}=1$ supergravity was explained, and the ${\cal W}$ term was written in \cite[(17.19)]{Freedman:2012zz}.

\subsection{Solution of the Lagrange multiplier constraints}
\label{ss:constraints}

Let us look at the term $\left[\Lambda  (X\nmult )^2\right] _F$
\begin{eqnarray}
e^{-1}{\cal L}_\Lambda &=&F^\Lambda\,(X\nmult )^2 +  \Lambda \left(2X\nmult F\nmult -\bar \chi\nmult  P_L\chi\nmult \right)\, -2\bar \chi^\Lambda P_L\chi\nmult X\nmult \nonumber\\
&&+\frac{1}{\sqrt{2}}\bar \psi_\mu \gamma^\mu \left(2\Lambda X\nmult P_L\chi\nmult + (X\nmult )^2P_L\chi^\Lambda \right)\nonumber\\
&&+\ft12 \bar \psi _{\mu }P_R \gamma ^{\mu \nu }\psi _{\nu }\Lambda (X\nmult )^2   +\hc
 \label{fullactionFformLambda}
\end{eqnarray}
The field equation of $\Lambda$ is
\begin{equation}
   2X\nmult F\nmult -\bar \chi\nmult  P_L\chi\nmult  +\sqrt{2}\bar \psi_\mu \gamma^\mu X\nmult P_L\chi\nmult +\ft12 \bar \psi _{\mu }P_R \gamma ^{\mu \nu }\psi _{\nu }(X\nmult )^2 =0\,.
 \label{fieldeqnLambda}
\end{equation}
This is solved as in the rigid case by
\begin{equation}
  X\nmult =\frac{\overline{\chi\nmult }P_L \chi\nmult  }{2F\nmult }\equiv \frac{\chi^2 }{2F}\,, \qquad F\nmult\equiv F \, , \qquad \chi\nmult \equiv \chi \ ,
 \label{Xnsolved2}
\end{equation}
since this kills all components of the chiral multiplet $(X\nmult )^2$.
It follows that the remaining equations for $\chi^\Lambda$ and $F^\Lambda$ are also satisfied.

\subsection{Details of $ [\ft12 \eta_{IJ}X^I\bar F^J]_F$ and $ [\mathcal{W}(X^I)]_F$ }
\label{ss:details}

Using \cite{Freedman:2012zz} as described above, the  first two terms of \rf{LinFformat} can be written as
\begin{eqnarray}
e^{-1}{\cal L}&=&\ft12\eta_{IJ}\left(F^I\bar F^J + X^I\bbox^C \bar X^J -\bar \chi^I P_L\slashed{\cal D}\chi^J\right)\nonumber\\
&& +{\cal W}_I F^I -\ft12{\cal  W}_{IJ}\bar \chi^{I}P_L\chi ^{J}\nonumber\\
  &&+\frac{1}{\sqrt{2}}\bar \psi_\mu \gamma
  ^\mu \left[\ft12\eta_{IJ}\left(P_L\chi^I \bar F^J + X^I\slashed{\cal D}P_R\chi^J\right)+{\cal  W}_I P_L \chi^I\right]\nonumber\\
&&+\ft12 \bar \psi _{\mu }P_R \gamma ^{\mu \nu }\psi _{\nu }\left(\ft12\eta_{IJ}X^I\bar F^J+{\cal W}  \right) +\hc
 \label{fullactionFform}
\end{eqnarray}
The superpotential and its first and second derivatives which we need in \rf{fullactionFform} are:
\begin{eqnarray}
 \mathcal{W}&=& a\left(\frac{X^0}{\sqrt{3}}\right)^3 + b\left(\frac{X^0}{\sqrt{3}}\right)^2 X\nmult \,, \nonumber\\
 \mathcal{W}_0 & = & 3 a\frac{(X^0)^2}{(\sqrt{3})^3} +\frac23bX^0 X\nmult \,,\qquad \mathcal{W}_{\onen }= \frac13b (X^0)^2\,,\nonumber\\
 \mathcal{W}_{00}&=& 6 a\frac{X^0}{(\sqrt{3})^3} +\frac23b X\nmult \,,\qquad \mathcal{W}_{0\onen }=\frac23 b X^0\,.
 \label{Wderiv}
\end{eqnarray}
This action can be written in the form  of Eqs. \rf{12XbarXD}, \rf{12tr0C0} in Sec. \ref{ss:derivation} (after noting that the Weyl connection  $b_\mu $ terms cancel).

The next major step is the elimination of auxiliary fields, but it is useful to first make some simplifications in our superconformal action. This will facilitate the derivation of  (\ref{actionSC}).  The simplifications are possible because we know that all the gauge connections recombine in covariant derivatives in order to make field equations supercovariant.  It  saves a lot of work to recognize this structure.
In particular, the
equation of motion for  $X^1$ should  be a conformally covariant equation modulo other field equations.  We start by writing the $\bar X^1$ field equation, before imposing the constraint:
\begin{eqnarray}
  e^{-1}\frac{\delta{\cal L}_1}{\delta \bar X\nmult }&=&\bbox^C X\nmult +\overline{{\cal W}}_{0\onen }\bar F^0-\ft12\overline{{\cal W}}_{00\onen }\bar \chi^{0}P_R\chi ^{0}\nonumber\\
  &&+\frac{1}{\sqrt{2}}\bar \psi_\mu\gamma^\mu\left[\slashed{\cal D}P_L\chi\nmult +\overline{{\cal W}}_{0\onen }P_R\chi^0+\frac{1}{\sqrt{2}}(F\nmult +\overline{{\cal W}}_{\onen })P_R\gamma^\nu\psi_\nu\right]\nonumber\\
  && -\ft12\bar \psi_\mu P_L\psi^\mu\left[ F\nmult +\overline{{\cal W}}_{\onen }\right]\,.
 \label{XL1fe}
\end{eqnarray}
Note that the expression in square brackets in the second line is the field equation of $P_R\chi\nmult $, while the one in the third line is the field equation of $\bar F\nmult $.
Writing out some covariant derivatives leads to further simplifications. One of these is that terms with $F^1$ all cancel. These simplifications lead to
\begin{eqnarray}
  e^{-1}\frac{\delta{\cal L}_1}{\delta \bar X\nmult }&=&\bbox'^C X\nmult +\overline{{\cal W}}_{0\onen }\bar F^0-\ft12\overline{{\cal W}}_{00\onen }\bar \chi^{0}P_R\chi ^{0}\nonumber\\
  &&+\frac{1}{\sqrt{2}}\bar \psi_\mu\gamma^\mu \overline{{\cal W}}_{0\onen }P_R\chi^0
 + \frac{1}{\sqrt{2}}\bar \psi_\mu\gamma^{\mu\nu}P_L{\cal D}_\nu'\chi\nmult
 + \frac{1}{2}\bar \psi_\mu\gamma^{\mu\nu}P_L\psi_\nu\overline{{\cal W}}_{\onen }\,.
  \label{XL1fesimpler}
\end{eqnarray}
One can see that this allows us to identify the terms $A_cX^1 + B_c$ in (\ref{actionSC}). The modified conformal derivatives that appear in \rf{XL1fesimpler} are given by
\begin{eqnarray}
 \bbox'^C X  & = & e^{a\mu}\left(\partial_\mu {\cal D}_a  X-2 b_\mu {\cal D}_a X +\chi_{\mu\,ab}{\cal
D}^b  X +2f_{\mu a}X+\rmi A_\mu {\cal D}_a  X  +\frac{1}{\sqrt{2}} \bar \phi _\mu \gamma
_aP_L\chi\right)\,,\nonumber\\
   {\cal D}_a  X &=& e _a^\mu \left(\partial _\mu   X-\,b_\mu X -\rmi \, A_\mu
   X -\frac{1}{\sqrt{2}}\bar \psi _\mu P_L\chi\right)\,, \nonumber\\
 P_L {\cal D}'_\mu\chi &=& P_L\left[ \left( \partial _\mu+\frac14\omega  _\mu {}^{bc}\gamma _{bc}
- \ft32 b_\mu   +\ft12\rmi A_\mu\right)\chi-\frac1{\sqrt{2}} \left(\slashed{\cal D} X \right)\psi _\mu  -\sqrt{2} X\phi _\mu \right]
\,.
\label{covderchiralmult}
\end{eqnarray}
As stated above the explicit $b_\mu$ terms cancel with those in the spin connection $\omega _\mu {}^{ab}=\omega _\mu {}^{ab}(e,b,\psi )$ and in $f^\mu_\mu$ (given in \cite[(16.26)]{Freedman:2012zz}¼.

\subsection{Gaussian integration of  auxiliary field \texorpdfstring{$F^0$}{F}}
\label{ss:elimauxf}

Since $F\nmult $ occurs  in the expression for $X\nmult $, we cannot use its field equation immediately.
But the other auxiliary fields, $F^0$ and $A_\mu $, are eliminated quite simply. We start with $F^0$;  its elimination preserves  the general structure of (\ref{actionSC}).

In order to eliminate the auxiliary field $F^0$, we first collect the terms in the action with $F^I$.
We write ${\cal L}_{1,F}$ as
\begin{equation}
{\cal L}_{1,F} =\eta_{IJ}\left(F^I+\eta^{IK}\overline{{\cal W}}_{\bar K}\right)\left(\bar F^J+\eta^{JL}{\cal W}_L\right)-
   {\cal W}_I \eta^{IJ}\overline{{\cal W}}_{\bar J}\,.
 \label{L1F}
\end{equation}
We eliminate $F^0$ and thus remain with
\begin{equation}
 {\cal L}_{1,F} \approx \left(F^1 + \overline{{\cal W}}_1\right)\left(\bar F^1 +{\cal W}_1\right)-{\cal W}_1 \overline{{\cal W}}_{\bar 1}+{\cal W}_0 \overline{{\cal W}}_{\bar 0}
 \,.
 \label{F0eliminate}
\end{equation}
Note that the term quadratic in ${\cal W}_0$ adds an additional term to the $A_c$ term, so that now after elimination of $F^0$ we have the following entries for \rf{actionSC}
\begin{eqnarray}
A_c&=&  \left[\partial _\mu \sqrt{g}g^{\mu\nu}\partial _\nu + \rmi\,e\, t_c^\mu \partial _\mu + \ft12\rmi\,\partial _\mu (e\,t_c^\mu) + e\,r_c\right] \ ,\\
r_c &=&r_0^c +{\cal W}_{01} \overline{{\cal W}}_{01}\nonumber\\
&=& -\ft16 R(\omega (e )) +\ft16\bar \psi _\mu \gamma ^{\mu \nu \rho } D^{(0)}_\nu \psi _\rho
  -A^aA_a -\ft16{\cal L}_{\rm SG,torsion}+\ft49|b X^0|^2 \,,\\
B_c&=& B^1_c+ \overline{W}_{01}\left[{\cal W}_0\right]_{X^1=0}\nonumber\\
& -&\frac12\overline{{\cal  W}}_{IJ1}\bar \chi^{I}P_R\chi ^{J}+\frac{1}{\sqrt{2}}\overline{{\cal  W}}_{I1}\bar \psi_\mu \gamma
  ^\mu P_R \chi^I+\frac12 \overline{{\cal W} }_1\bar \psi _{\mu }P_L \gamma ^{\mu \nu }\psi _{\nu }\nonumber\\
  &=& \frac{1}{\sqrt{2}}\left[-e^{-1}\partial _\mu \left(e\bar \psi _\nu \gamma ^\mu \gamma ^\nu P_L  \chi^1\right)-\frac23\bar \chi^1 P_L \gamma ^{\mu \nu }D _\mu \psi _\nu +\rmi A^\mu \bar \psi _\mu P_L\chi^1\right]\nonumber\\
& +&\frac{\bar b}{3}
\Bigl(2 \frac{1}{\sqrt{3}}a (X^0)^2\bar X^0 -{\bar\chi}^0P_R\chi^0 + \sqrt{2}{\bar\psi}\cdot\gamma P_R \chi^0 \bar X^0 +
\frac{1}{2}(\bar X^0)^2 {\bar\psi}_\mu\gamma^{\mu\nu}P_L\psi_\nu\Bigr).
\end{eqnarray}
\begin{eqnarray}
 C_c & = & -e^{-1}\bar X^0  \left[\partial _\mu \sqrt{g}g^{\mu\nu}\partial _\nu + \rmi\,e\, t_c^\mu \partial _\mu + \frac12\rmi\,\partial _\mu (e\,t_c^\mu) + e\,r_0^c\right] X^0 \nonumber\\
   &    -&\bar X^0B^0_c -X^0\bar B^0_c + C_0 + \left[{\cal W}_0\right]_{X^1=0}\left[\overline{{\cal W}}_0\right]_{\bar X^1=0}\nonumber\\
   & +&\left[ \left(-\frac12{\cal  W}_{IJ}\bar \chi^{I}P_L\chi ^{J}+\frac{1}{\sqrt{2}}\bar \psi_\mu \gamma
  ^\mu {\cal  W}_I P_L \chi^I+\frac12 \bar \psi _{\mu }P_R \gamma ^{\mu \nu }\psi _{\nu }\,{\cal W}\right)_{X^1=0} +\hc\right]
 \label{Cversie1}
\end{eqnarray}

\subsection{Gaussian integration of  auxiliary field \texorpdfstring{$A_\mu $}{A}.}
\label{ss:elimAmu}

Then we turn to the elimination of $A_\mu $. We write as in
 \cite[(17.21)]{Freedman:2012zz}
\begin{eqnarray}
e^{-1}\frac{\delta {\cal L}_1}{\delta A^\mu}
&=&\rmi\left[({\cal D} _\mu X^I)\eta_{IJ}\bar X^{\bar J}-\hc\right]+\frac{1}{2}\rmi\eta_{IJ}\bar \chi ^IP_L\gamma _\mu \chi^{\bar J}\nonumber\\
&=&2 A_\mu X^I\eta_{IJ}\bar X^J + \rmi\left[\left(\partial _\mu X^I+\frac{1}{\sqrt{2}}\bar \psi_\mu
P_L\chi^I\right)\eta_{IJ}\bar X^{\bar J}-\hc\right]\nonumber\\
&&+\frac{1}{2}\rmi\eta_{IJ}\bar \chi ^IP_L\gamma _\mu \chi^{\bar J}\,.
\end{eqnarray}
where
$
  \frac{1}{N}= -\frac{1}{X^0\bar X^0}-\frac{ X^1\bar X^1}{(X^0\bar X^0)^2}
 $.
The solution for $A_\mu $ is
\begin{eqnarray}
  A_\mu &=& {\cal A}_\mu +{\cal A}_\mu ^{\rm F}\,,\nonumber\\
  && {\cal A}_\mu= \rmi\frac{1}{2N}\eta _{IJ}(X^I\partial _\mu \bar X^J -\bar X^I\partial _\mu X^J)={\cal A}_\mu ^0+{\cal A}_\mu ^1 \,,\nonumber\\
  &&\phantom{.}\qquad {\cal A}_\mu ^0= \rmi\frac{1}{2N}(-X^0\partial _\mu \bar X^0 +\bar X^0\partial _\mu X^0)\,,\nonumber\\
  &&\phantom{.}\qquad {\cal A}_\mu ^1= \rmi\frac{1}{2X^0\bar X^0}(-X^1\partial _\mu \bar X^1 +\bar X^1\partial _\mu X^1)\,,\nonumber\\
  && {\cal A}_\mu ^{\rm F}=\frac{1}{4N}\rmi\eta _{IJ}\left[\sqrt{2}\bar \psi _\mu(P_L\chi ^J \bar X^I-P_R\chi ^J X^I) +\bar \chi ^IP_L\gamma_\mu \chi ^J\right]={\cal A}_\mu ^{{\rm F}0}+{\cal A}_\mu ^{{\rm F}1}\,,\nonumber\\
  &&\phantom{.}\qquad {\cal A}_\mu ^{{\rm F}0}= -\frac{1}{4N}\rmi\left[\sqrt{2}\bar \psi _\mu(P_L\chi ^0\bar X^0 -P_R\chi ^0 X^0) +\bar \chi ^0P_L\gamma_\mu \chi ^0\right]\,,\nonumber\\
  &&\phantom{.}\qquad {\cal A}_\mu ^{{\rm F}1}= -\frac{1}{4X^0\bar X^0}\rmi\left[\sqrt{2}\bar \psi _\mu(P_L\chi ^1\bar X^1 -P_R\chi ^1 X^1) +\bar \chi ^1P_L\gamma_\mu \chi ^1\right]\,.
 \label{Amusoln}
\end{eqnarray}
The terms ${\cal A}_\mu ^0$ and ${\cal A}_\mu ^{{\rm F}0}$ vanish after gauge-fixing (\ref{sconfgf}), and the on-shell value of ${\cal A}_\mu $ simplifies to
\begin{eqnarray}
A_\mu= \rmi\frac{\kappa^2}{6}[ (\bar X\partial _\mu X -X\partial _\mu \bar X )  -\frac{1}{2}\left[\sqrt{2}\bar \psi _\mu(P_L\chi \bar X -P_R  \chi  X) +\bar \chi P_L\gamma_\mu \chi \right]]\,.
 \label{AmuA}\end{eqnarray}
 After elimination, the entire effect of $A_\mu$ resides in
 a contribution to the Lagrangian %the quadratic term in $r_0^c$ in \rf{12tr0C0} which becomes
 \begin{equation}
 {\cal L}_A=e\,N A^\mu A_\mu = \frac{\kappa^2}{24}  \chi^2\bar \chi^2 \ .
 \end{equation}
This simple form arises only from the last term of \rf{AmuA} after Fierz rearrangement. It contributes to the expression $C$ in (\ref{actionS}), which is
\begin{eqnarray}
 C&=& \frac{1}{2\kappa ^2} \left[ R(\omega (e )) -\bar \psi _\mu \gamma ^{\mu \nu \rho } D^{(0)}_\nu \psi _\rho
  +{\cal L}_{\rm SG,torsion} \right] +3\frac{m^2}{\kappa ^2} -f ^2 \nonumber\\
&&+\frac{1}{\sqrt{2}}f \bar \psi_\mu \gamma
  ^\mu   \chi+\frac{m}{2\kappa ^2}\bar \psi _{\mu } \gamma ^{\mu \nu }\psi _{\nu } +\frac{\kappa ^2}{24}\chi ^2\bar \chi ^2\nonumber\\
 &&-\ft12\bar \chi  \slashed{D}^{(0)}\chi -\ft{1}{32}\rmi\,e^{-1}\varepsilon ^{\mu \nu \rho \sigma }\bar \psi _\mu \gamma _\nu \psi _\rho \bar \chi \gamma _*\gamma _\sigma \chi-\ft12\bar \psi _\mu P_R\chi \bar \psi ^\mu P_L\chi\,.
 \label{formC}
\end{eqnarray}

\subsection{Non-Gaussian integration of auxiliary field \texorpdfstring{$F$}{F}}
\label{app:solnF}
Here we give the detailed derivation of results in the last part of Sec. \ref{ss:derivation}. We start with the action \rf{actionS}
 where
$X= \frac{\chi^2}{ 2 F} $ and $\bar X= \frac{\bar \chi^2}{ 2 \bar F}$.  Then we solve for the fields $F$ and $\bar F$ using their algebraic equations of motion.
The field equation for $F$ is
\begin{equation}
\frac{\delta{\cal L}(X,F)}{\partial \bar F}-\frac{\bar X}{\bar F}\frac{\delta{\cal L}(X,F)}{\partial \bar X}=0\  \ \longrightarrow \ \
F + f - \frac{\bar X}{\bar F} \left( A\, X + B\right)=0\,.
\label{Ffieldeqn}
\end{equation}
This implies  that
\begin{equation}
  F = -f+{\cal O}(\bar \chi^2)\,,\qquad \bar F = -\bar f+{\cal O}(\chi^2)\,,
 \label{Ffshort}
\end{equation}
where e.g. ${\cal O}(\bar \chi^2)$ means that the correction terms are proportional to an undifferentiated $\bar \chi^2$.
The complete expression is
\begin{equation}
 F = -  f \left[ 1-   \frac{\bar X}{ f \bar F} \left( A\,  X +  B\right)\right]\,.
 \label{FexpressionX}
\end{equation}
Since $\bar X$ is nilpotent, we have also
\begin{equation}
F^{-1} = -\frac{1}{f} \left[ 1+   \frac{\bar X}{ f \bar F} \left(A\, X + B\right)\right]\,.
\label{barF-1}
\end{equation}
This allows to write the following expression for $X$
\begin{eqnarray}
X &=& \frac{\chi^2}{ 2}F^{-1}  =  - \frac{\chi^2}{ 2 f}  \left[ 1+   \frac{\bar X}{ f \bar F} \left(A\, X + B\right)\right]\nonumber\\
&=&- \frac{\chi^2}{ 2 f}  \left[ 1+   \frac{\bar \chi^2}{ 2f \bar f^2} \left(A\, \frac{\chi^2}{2F} + B\right)\right]\nonumber\\
&=&- \frac{\chi^2}{ 2 f} \left[ 1-   \frac{\bar \chi^2}{  2f \bar f^2} \left(A\, \frac{\chi^2}{2f} - B\right)\right]\,,
\end{eqnarray}
where the second line is obtained using (\ref{Ffshort}) and for the third line we observe that the two derivatives in $A$ must both act on $\chi^2$ in order not to be killed by the overall factor $\chi^2$.\footnote{In fact, we could move $f$ outside of the $A$ operator, and even replace the $A$ by only its part $\bbox$, but this is not convenient for what follows below.}
We define now for convenience
\begin{equation}
  {\cal A}= \frac{\bar \chi^2}{ 2f \bar f^2} \left(A\, \frac{\chi^2}{2f} - B\right)
  \,.
 \label{defcalA}
\end{equation}

The quantity ${\cal A}$ is thus fully determined by the functions $A$, $B$ and $f$ that appear in the action and the fermionic composite scalar $\chi^2$ (and their complex conjugates). The dependent field $X$ is
$
  X= - \frac{\chi^2}{ 2 f} (1-{\cal A})\,.
 $  In order to find $F$ we have to consider
\begin{eqnarray}
  \bar X [AX+B]&=& (1-\bar {\cal A}) \frac{\bar \chi^2}{ 2 \bar f}\left[A\left(\frac{\chi^2}{2 f} (1-{\cal A})\right)-B\right]\nonumber\\
  &=&(1-\bar {\cal A})\left[f\bar f{\cal A}- \frac{\bar \chi^2}{ 2 \bar f}A\left(\frac{\chi^2}{ 2 f}{\cal A}\right)\right]\,.
 \label{XAXB}
\end{eqnarray}
In the last term, the $A$ should fully act on the leading factor of ${\cal A}$ in (\ref{defcalA}) in order that this factor does not clash with the leading $\bar \chi^2$. It should also fully act as the $\bbox$ factor, which means that we can write $A$ also as $\bar A$ in order to get the following elegant equation:
\begin{eqnarray}
  \frac{\bar \chi^2}{ 2 \bar f}A\left(\frac{\chi^2}{2 f}{\cal A}\right)&=&\frac{\bar \chi^2\chi^2}{ 4 f\bar f}\left(\bar A\frac{\bar \chi^2}{  2f \bar f^2}\right)\left(A\, \frac{\chi^2}{2f} - B\right)\nonumber\\
  &=&{\cal A}\frac{\chi^2}{2f}\bar A\frac{\bar \chi^2}{2\bar f}={\cal A}\left(f\bar f\bar{{\cal A}}+\frac{\chi^2}{2f}\bar B\right)\,.
 \label{AoncalA}
\end{eqnarray}
Introducing this in (\ref{XAXB}) and using the nilpotency of ${\cal A}$ and $\bar {\cal A}={\cal O}(\chi^2)$ gives
\begin{equation}
 \bar X [AX+B] ={\cal A}f\bar f\left[1-2\bar{{\cal A}}-\frac{\chi^2}{2f^2\bar f}\bar B\right]\,.
 \label{resultXAXB}
\end{equation}

We find therefore with (\ref{FexpressionX})
\begin{equation}
  F = -  f\left[1-   \frac{\bar f}{   \bar F}{\cal A} \left(1-2\bar{{\cal A}}-\frac{\chi^2}{2f^2\bar f}\bar B\right)\right]\,.
 \label{Fplusf}
\end{equation}
This implies e.g.
\begin{equation}
  -\frac{\bar f}{\bar F}= 1-\bar {\cal A}+{\cal O}(\chi^2\bar \chi^2)\,,
 \label{ccfF}
\end{equation}
which gives as final expression for $F$:
\begin{equation}
  F= -  f\left[1+{\cal A} \left(1-3\bar{{\cal A}}-\frac{\chi^2}{2f^2\bar f}\bar B\right)\right]\,.
 \label{finalF2}
\end{equation}
Due to the orders of nilpotent quantities, we also obtain
\begin{equation}
  (F+f)(\bar F+\bar f)= f\bar f{\cal A}\bar {\cal A}\,.
 \label{Ff2}
\end{equation}
Also, the other quantity that appears in  the action simplifies:
\begin{equation}
  \bar X(AX+B)+X\bar B = {\cal A}f\bar f\left[1-2\bar{{\cal A}}\right]-\frac{\chi^2}{2f}\bar B\,.
 \label{termLagrangianA}
\end{equation}
Observe that
\begin{equation}
  f\bar f{\cal A}= \frac{\bar  \chi^2}{2\bar f}A \frac{\chi^2}{2 f} -\frac{\bar \chi^2}{2\bar f}B\,,
 \label{realtermA}
\end{equation}
where the first term is real up to a total derivative, such that the expression (\ref{termLagrangianA}) leads to a real action.
We can write the whole Lagrangian (\ref{actionS}) as
\begin{eqnarray}
 e^{-1} {\cal L}&=& f\bar f\left(-1+{\cal A}-{\cal A}\bar {\cal A}\right)-\frac{\chi^2}{2f}\bar B+C\nonumber\\
  &=& -f\bar f+\frac{\bar  \chi^2}{2\bar f}A \frac{\chi^2}{2 f} -\left(\frac{ \chi^2}{2 f} \bar B+\frac{\bar \chi^2}{2\bar f} B\right)
  +C\nonumber\\
  &&- \frac{\chi^2\bar \chi^2}{16(f\bar f)^2}\left(f^{-1}\bbox\chi^2-2B\right)\left(\bar f^{-1}\bbox\bar \chi^2-2\bar B\right)  \,.
 \label{finalactionfreal}
\end{eqnarray}

%%%%%%%%%%%%%%%%%%%%%%%%%%%%%%%%

%\newpage
%%%%%%%%%%%%%%%%%%%%%%%%%%%
%\appendix
%\section{Notation}
%%%%%%%%%%%%%%%%%%%%%%%%%%%%%%%%%%%%%%%%%%%%%%%%%%%%%%%
%\bibliography{supergravity}
%%%Included for WinEdt Gather Purpose (do not remove the comment line below:
%%             %input "C:\localtexmf\bibtex\bib\*.bib"
%%             %input "C:\Program Files\MiKTeX\texmf\bibtex\bib\*.bib"
%%             %input "C:\localtexmf\Bibtex\bib\*.bib"
%\bibliographystyle{toine}
\end{document}